# Soft Computing Tools To Predict Varied Weight Components, Material and Tribological Properties of Al2219-$B_4$C-Gr

Maitreyi Chatterjee[1] and Biplab Chatterjee[2]
[1]Computer Science, Cornell University , Ithaca, USA
[2]Ground Handling, AI Airport Services Ltd, Kolkata, India

**ABSTRACT**
Soft computing tools emerged as most reliable alternatives of traditional regression and statistical methods. In recent times, these tools can predict the optimum material compositions, mechanical and tribological properties of composite materials accurately without much experiment or even without experiment. In the present study, soft computing tools like fuzzy logic, Decision tree, genetic algorithms are employed to predict the reinforcement weight percentage of $B_4$C(Boron Carbide) and Graphite(Gr) along with Aluminum (matrix material) weight percentage for Al2219 with $B_4$C and graphite. The optimized material and tribological properties of Al2219 were also predicted using NSGA II genetic algorithms for multi-objective optimization. It is found that the predictions are at par with earlier ANN (artificial neural network) studies and experimental findings. It can be inferred that inclusion $B_4$C has more impact on enhancement of mechanical properties as well as wear strength compared to Gr.
**Keywords:** *Fuzzy logic, Decision Tree, Genetic algorithm, Multiobjective Optimisation*

**NOMENCLATURE**

| | |
|---|---|
| COF | Coefficient of friction (dimensionless) |
| $E$ | Elastic Modulus (GPa) |
| $E_m$ | Elastic Modulus of matrix material (GPa) |
| $E_g$ | Elastic Modulus of Graphite(GPa) |
| $E_b$ | Elastic Modulus of Boron carbide(GPa) |
| $W$ | Wear Rate |
| $\sigma$ | Ultimate Tensile Strength(MPa) |

## 1. INTRODUCTION

Computational intelligence plays a pivotal role in component selection and to predict it's behaviour in composite materials with limited experimentation or even without any experiment in today's artificial intelligence and machine learning[12][13][14] era. Sarma et al. [1] presented a review paper about the trend and progress of exploring the use of machine learning algorithm for polymer composite. Vinoth and Datta [2] applied ANN and genetic algorithm [3] for the development of multiple reinforcement component and it's optimisation to get the material properties of ultrahigh molecular weight polyethylene composites [UHMWPE]. They made data bank from 153 datasets from the existing literature. In similar fashion, Olfatbakhsh, and Milani [4] presented different machine learning methods to predict the material property in case of woven fibre reinforced fabrics. Several researchers are concentrating on ANN for data analysis in wide areas. Tsao and Hocheng [5] evaluated thrust force and surface roughness using neural network. Park and Rhee [6] made ANN as a tool to predict the UTS of the welded joint of Al5182 and Al5356 and also applied genetic algorithm for the optimisation of welding parameters. Radha et al. [7] have utilised Back Propagation Neural Network [BPNN], Radial Basis Neural Network [RBNN], Fuzzy Logic [FL], Decision Tree like Neural Network/computational intelligence tools to study the stress, strain, % of elongation and impact energy along with the tribological properties of Al 2219 – ($B_4$C + H-BN) for different load and $B_4$C content. In this study, the aim is to compare the mechanical and tribological properties of Aluminium metal matrix composites reinforced with $B_4$C and graphite. With the help of Fuzzy logic (FL), the present study developed crisp Al2219 +( $B_4$C – Gr) material composition, studied the optimisation of mechanical properties and tribological properties through NSGA II to get the Pareto optimal front [8]. Also, tribological behaviour is studied through decision tree.

## 2. COMPUTATIONAL METHODS

In this study, different computational methodologies are used, namely Fuzzy logic, decision tree, optimisation tools like NSGA II for multi objective optimisation for material properties E, σ, tribological properties W and COF. The methodologies applied in this paper are briefly described in the following sub-sections.

**2.1 Fuzzy logic**

Fuzzy logic based on reasoning: It is rule based framework, not everything in black and white or like true/false. In fuzzy logic, parameter ranges in between 0 and 1. Fuzzy interference system(FIS) converts the crisp data into fuzzy variables like very low, low, medium, high, very high. We considered E and σ in low, medium and high as Fuzzy variables.The process is termed as fuzzification. Then it applies the if-then rules depending on fuzzy values. Interference engine then evaluates the outcome using fuzzy logic. The last stage is the defuzzification where it returns the variables in terms of crisp data. Real world uses involves washing machine (determines the number of wash cycles depending on dirt level), autonomous vehicle (Smooth control over steering, braking) etc.

### 2.2 Decision Tree

Decision tree is a supervised learning used for classification and regression purpose which creates tree like structure wherein nodes denote conditions on features and leaves indicates final prediction (class labels or values). It is basically a flow chart initiated from root (top most node) and it branches into conditions on feature values that ends in outcomes. Classification rule executed through the path from root to leaf. It is exclusively used in the quality control decision in manufacturing, medical diagnosis for a particular deceases. For approval in loan or credit card in finance, the application of decision tree for issuing a credit card is as follows:

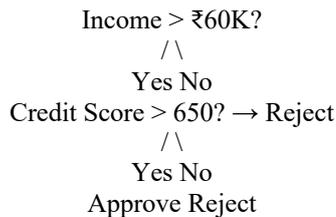

```
         Income > ₹60K?
           /    \
          Yes    No
   Credit Score > 650?  → Reject
         /    \
        Yes    No
      Approve  Reject
```

For calculating COF, we have considered different values of load(sample size,20) and varied $B_4C$, wheras for computing wear different sliding velocity used with load.

### 2.3 Genetic Algorithm

Genetic algorithm(GA) is an optimisation procedure and search technique through natural selection. It's Darwin thinking applied in data analysis. It comprises of three phases named as selection, cross over and mutation. Selection is the process to choose the fittest solutions (individuals) as parent for next generation, cross over is the combination of pairs of both parents that give a solution which inherit distinguish quality or characteristics from both. Mutation is the process to impose small random changes for diversity and to prevent premature convergence. The process repeats and after consecutive generations, optimal solution is evolved. The optimal solution or optimisation of a objective function with or without the constraint in GA can be defined when i) no significant changes per generation ii) maximum number of generations explored iii) desired fitness, if known in advance, attended. In real world, the application of GA are used in optimal design of control system and structural design in engineering, feature selection & neural architecture search in artificial intelligence /machine learning and for path finding/ motion planning in Robotics.

### 2.4 Multi Objective Optimisation through GA

In a single objective optimisation problem, we are getting one best solution. In case of multiple objective, the aim is to search for Pareto optimality. In Pareto-optimality, it is not possible to improve one objective without worsening at least one other objective within the feasible range of Pareto sets. In our study of multi objective function, in some cases even for multiple conflicting objective function (E, σ)we have studied for a set of trade off solutions called the Pareto front. Applications in real world are the balancing of weight and strength of materials in engineering, maximum accuracy and minimum complexity in machine learning, minimum energy and maximum speed in Robotics.

### 3. RESULTS AND DISCUSSIONS

Theoretically the material properties of composite materials are computed by mixture rule [9]. We have considered $E_m$=80 GPa, $E_b$=470 GPa, $E_g$=15 GPa and $σ_b$=569 MPa, $σ_g$=34 MPa from [9]. For percentage variation, Aluminium is varied from 75 to 100, $B_4C$ from 0 to 15 and Graphite 0 to 10 [10]. The formula for mixture rules are used as objective function in applying NSGA II in a programme in Python 3.11.13.

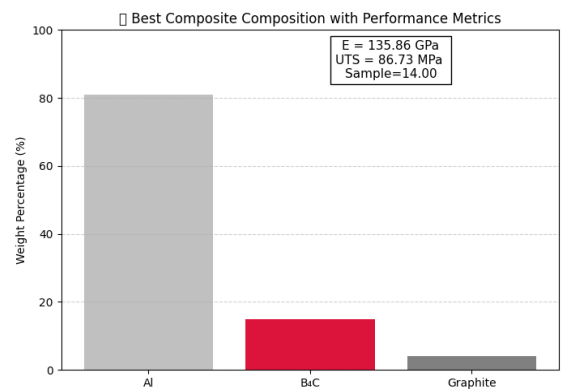

**Figure: 1 Best Composite Composition with Performance Metrics**

Figure 1 presents the best composite composition with Pareto front value of elastic modulus and ultimate tensile Strength. The best trade-off composite: Al = 80.93%, $B_4C$ = 15.00%, Gr = 4.07%, E=135.86 GPa, σ=86.73 MPa. This complies with the result of Nagaraju et al. [10] where ANN results show 100% contribution of $B_4C$ and less than 50% Graphite is necessary for higher Young's modulus. Similarly, ANN predicted 100.0% $B_4C$ and less than 55% Gr is necessary to achieve a higher UTS.

**Table 1: Crisp data through Fuzzy logic**

| E(GPa) | σ (MPa) | AL% | $B_4C$% | Gr% |
|---|---|---|---|---|
| 60 | 90 | 80 | 10 | 10 |
| 62.54 | 89.43 | 82 | 10 | 8 |
| 65.01 | 89.83 | 84 | 10 | 6 |
| 66.56 | 89.83 | 86 | 10 | 4 |
| 66.29 | 89.43 | 88 | 10 | 2 |
| 62.54 | 83.25 | 92 | 6 | 2 |
| 61.14 | 78.5 | 94 | 4 | 2 |
| 59.17 | 76.33 | 96 | 2 | 2 |
| 56.33 | 76.33 | 98 | 2 | 0 |
| 53.33 | 73.33 | 100 | 0 | 0 |
| 66.29 | 76.33 | 88 | 2 | 10 |
| 66.56 | 89.83 | 86 | 4 | 10 |
| 65.01 | 82.97 | 84 | 6 | 10 |
| 62.54 | 86.48 | 82 | 8 | 10 |

Fuzzy logic is frame work for reasoning under uncertainty. It is used in machine control system. Fuzzy logic measures the value as low, medium, high and ranges from 0 to 1 rather than the crisp [7]. Table 1 is made using FIS (Fuzzy interface system) from the logic of raw experimental data. FIS evaluates the input and converts it into crisp output using defuzzification. Logic is followed from Nagaraju et al. [10] and Sharath et al. [11]. Reinforcement of graphite enhances the ultimate tensile strength than unreinforced alloy. $B_4C$ is mostly used in high temperature system. It is also reported that inclusion of $B_4C$ enhances UTS. We have considered the variation of weight percentage of Al from 80 to 100 percent, graphite and $B_4C$ from 0 to 10 percent. Total variation of particulate ($B_4C$ and Gr) upto 20 percent [11] but intentionally varied weight percentage equally to establish the contribution of each particulate in Pareto optimisation and generate fourteen(sample size)crisp data for studying the optimisation of mechanical properties. Elastic modulus (for fuzzification) is considered in the range from 50(low) to 70(high) GPa and ultimate tensile strength (for fuzzification) is considered in the range from 70(low) to 95(high)MPa. Figure 2 is the plot of objective space along with the weight percentage of particulates from Fuzzy generated data. It is observed from the figure that varying the weight percentage upto certain limit both the mechanical properties significantly increase but the values of E and σ are not maximum at the total 20 weight percent of particulate as we have considered during the formation of crisp values using fuzzy logic. Higher contents of particulates create irregularity of mechanical properties, also observed in earlier studies [2]. The Pareto front or optimal values of E and σ are 66.56 GPa and 89.83 MPa respectively.

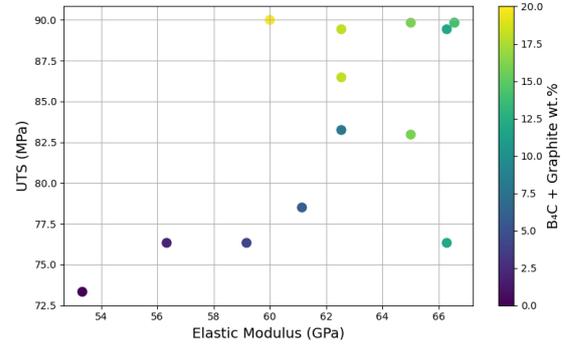

**Figure: 2 Objective Space from Fuzzy logic**

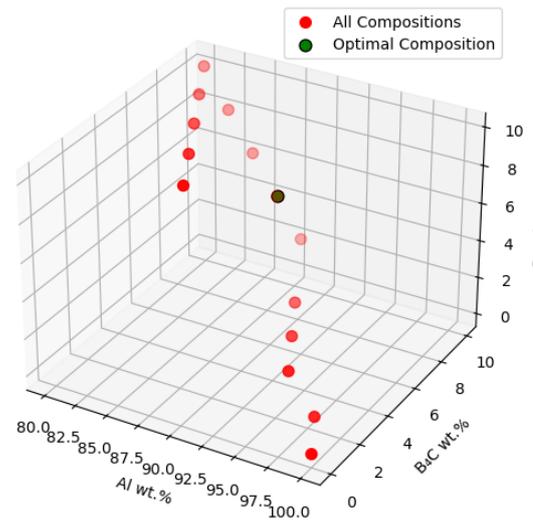

**Figure: 3 Optimal Composition at Design Space**

Figure 3 presents the design space of Aluminium, $B_4C$ and Gr weight percentage. The best weight percentage for the Pareto front of the mechanical properties are Al = 86.0%, $B_4C$=10.0% and Graphite = 4.0%. These values are at par with the findings reported in Ref. [10] as discussed earlier. Figure 4 and 5 present the decision tree with different load and $B_4C$ to predict COF and varying sliding speed with different loads for wear. [7] and Sarath et al. [11]. Radha et al. [7] reported if $B_4C$% is less than 3% then the CoF value is predicted as 0.51733. Similarly to predict CoF value as 0.44189, the input parameter load is greater than or equal to 6 and $B_4C$% is greater than 5. This conforms with our results of micro composites. Similarly our results for wear under 400 meter distance with varying load and sliding speed condition complies with findings reported in Ref.[11]. Figure 6 presents objective space of nano composite as per the sample data from Ref. [7]. From the NSGA II analysis, for nano composite the design variables load, Al%, B$_4$C% are 4, 0.4% and 8% which results the best minimum CoF and wear rate as 0.25 and 0.56 respectively, confirms with[7].

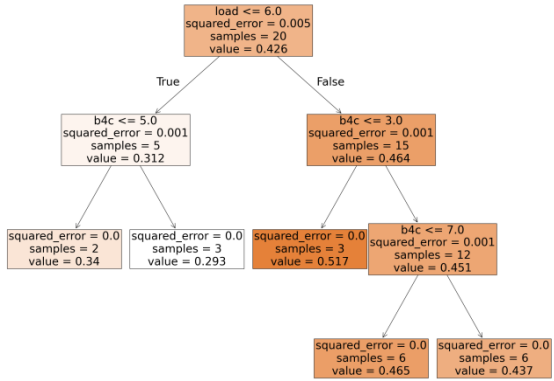

**Figure: 4 Decision Tree for COF of composite**

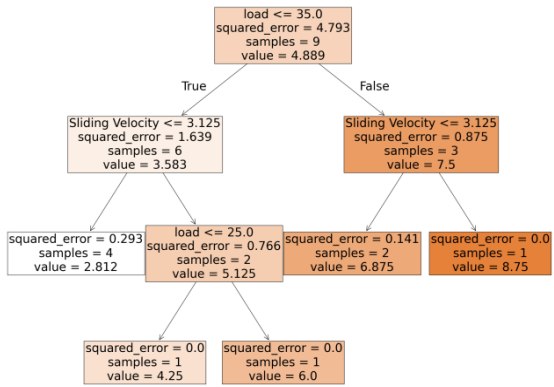

**Figure: 5 Decision Tree for Wear for S1 sample [Ref. 11]**

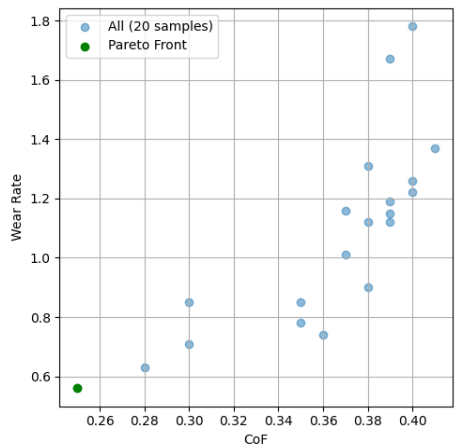

**Figure: 6 Objective Space of Nano composites**

## 4. CONCLUSIONS

The developed soft computing tools predict the reinforced particulate weight percentage, material and tribological properties of Al2219-B$_4$C-Gr composite. Different supervised and unsupervised computational intelligence tools serve to predict various parameters of the material. Fuzzy logic was used to predict crisp values of Al2219-B$_4$C-Gr components. Multi-objective optimizations were performed based on NSGA II genetic algorithms. The predictions had provided the accurate results with previous findings. It can be concluded that for the improvement of different properties of the materials to serve the growing demands in today's life, soft computing tools are the prerequisites in order to minimize the time requirement for experiment, space for experiment and the cost for the development of new features in the materials. Present results confirm that B$_4$C has more impact than Gr in enhancing mechanical properties (around 60 %) and wear strength. Decision tree of a sample (Al=90%, B$_4$C=5% and Gr=5% ) with 400 m sliding distance with varying load and sliding speed suggest that simultaneous increase of load and sliding speed increases the wear rate. Future work can explore the experimental validation of the composite materials.